\documentclass[aps,prl,twocolumn,showpacs]{revtex4-1}
\usepackage{amsmath}
\usepackage[dvips,xdvi]{graphicx}
\usepackage{amssymb}
\usepackage{epsfig}

\begin{document}
\title{Mapping the phase diagram of spinor condensates via adiabatic quantum phase transitions}
\author{J. Jiang}
\author{L. Zhao}
\author{M. Webb}
\author{Y. Liu}
\email{yingmei.liu@okstate.edu} \affiliation{Department of
Physics, Oklahoma State University, Stillwater, OK 74078}
\date{\today}

\begin{abstract}
We experimentally study two quantum phase transitions in a sodium
spinor condensate immersed in a microwave dressing field. We also
demonstrate that many previously unexplored regions in the phase
diagram of spinor condensates can be investigated by adiabatically
tuning the microwave field across one of the two quantum phase
transitions. This method overcomes two major experimental
challenges associated with some widely used methods, and is
applicable to other atomic species. Agreements between our data
and the mean-field theory for spinor Bose gases are also
discussed.
\end{abstract}


\maketitle

A spinor Bose-Einstein condensate (BEC) is a multi-component BEC
with an additional spin degree of freedom, which has provided
exciting opportunities to study quantum magnetism, superfluidity,
strong correlations, spin-squeezing, and massive
entanglement~\cite{StamperKurnRMP, Ueda, Duan2013, Chapman2013,
Chapman2012}. The interesting interactions in spinor BECs are
interconversions among multiple spin states and magnetic field
interactions (or microwave dressing field interactions)
characterized by $q_{\rm net}$, the net quadratic Zeeman energy.
The interplay of these interactions leads to oscillations among
multiple spin populations, which has been experimentally confirmed
in $F$=1 $^{23}$Na spinor BECs~\cite{JiangBEC,ZhaoUwave,faraday,fluctuation,Raman2011,black,Gerbier2012},
and in both $F$=1 and $F$=2 $^{87}$Rb spinor
condensates~\cite{Chapman2005,bloch,f2Hirano,f2sengstock1,f2sengstock2}.

Several groups demonstrated the mean-field (MF) ground states of
spinor BECs by holding BECs in a fixed magnetic field and letting
spin population oscillations damp out over a few
seconds~\cite{black,faraday,Gerbier2012,fluctuation}. The required
damping time, determined by energy dissipation, may in some cases
exceed the BEC lifetime. The exact mechanism involved in energy
dissipation requires further study, although it has been shown
that energy dissipates much faster in high magnetic
fields~\cite{fluctuation}. For $F$=1 BECs, a magnetic field
introduces only a positive $q_{\rm net}$, while a microwave field
has a distinct advantage since it can induce both positive and
negative $q_{\rm net}$~\cite{ZhaoUwave,uwaveBloch,Stamper-Kurn2009,StamperKurnRMP,Raman2011}.
As shown in Ref.~\cite{ZhaoUwave}, the same physics model explains
spin-mixing dynamics observed in both microwave fields and
magnetic fields. One would assume that, if given a long enough
exposure to a microwave field, a spinor BEC could also reach its
MF ground states. However, experimental studies on ground states
of spinor BECs in microwave fields have proven to be very
difficult, since these fields are created by near-resonant
microwave pulses. Two major experimental challenges associated
with microwave fields are atom losses and variations in
magnetization $m$. Microwave-induced changes in both $m$ and the
atom number $N$ can be detrimental, especially when a spinor BEC
is exposed to a large microwave field for a prolonged
time~\cite{ZhaoUwave,Raman2011}. As a result, the phase diagram of
$F$=1 BECs has not been well explored in the $q_{\rm net}\leq 0$
region, where applying microwave fields may be necessary.

In this paper, we demonstrate a new method to overcome the
aforementioned experimental challenges and report the observation
of two quantum phase transitions in a spinor BEC. In this method,
we quickly prepare an initial equilibrium state at a very high
magnetic field to minimize the damping time for spin population
oscillations and prevent unnecessary exposure to microwave pulses.
Equilibrium states at a desired $q_{\rm net}$ are then created by
adiabatically sweeping an additional microwave field. Using this method, we
are able to investigate many previously unexplored regions in the
phase diagram of antiferromagnetic spinor BECs and observe three
distinct quantum phases. Similarly to Ref.~\cite{Ueda,
StamperKurnRMP,Gerbier2012}, we define three phases in the MF
ground states based on $\rho_0$, the fractional population of the
$|F=1, m_F =0 \rangle$ state: $\rho_0=1$, $\rho_0=0$, and
$0<\rho_0<1$ respectively represent a longitudinal polar phase, an
antiferromagnetic (AFM) phase, and a broken-axisymmetry (BA)
phase. We observe two quantum phase transitions: one is between a
longitudinal polar phase and a BA phase at a fixed positive
$q_{\rm net}$, and the other is an AFM-BA phase transition at a given
$m$. We also calculate the energy gap between the ground states
and the first excited states in a spinor BEC, which provides an
explanation for the feasibility of this new method. In addition,
spin domains and spatial modes are not observed in our system, and
our data can be well fit by predictions of the single spatial-mode
approximation (SMA).
\begin{figure}[t]
\includegraphics[width=85mm]{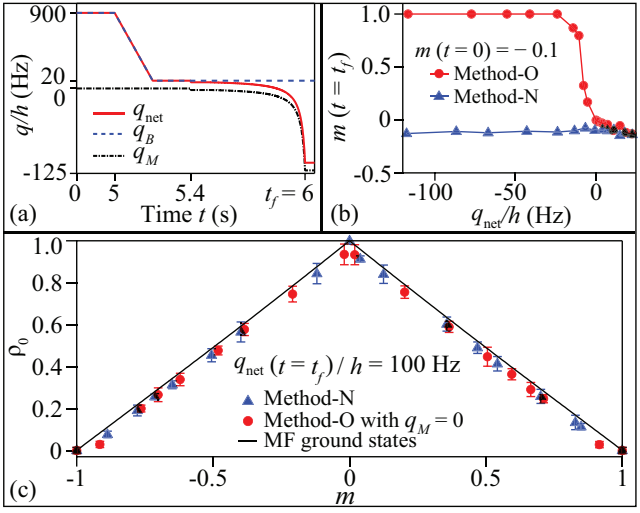}
\caption{(color online) (a). A typical experimental sequence of
Method-N, which is our new method to create equilibrium states via
adiabatically sweeping a microwave field. In this paper $-150~{\rm
Hz}\leq q_{\rm net}(t=t_f)/h\leq$~150~Hz. All axes are not to
scale. (b). $m$ as a function of $q_{\rm net}$ at $t=t_f$ in the
two methods starting from the same initial state, i.e.,
$m(t=0)=-0.1$. Note that $t_f$ for Method-O in this panel is only
1~s, which is much shorter than the typical hold time for creating
equilibrium states. (c). $\rho_0$ as a function of $m$ at $q_{\rm
net}(t=t_f)/h=100$~Hz in equilibrium states created by the two
methods. In this panel, Method-O prepares equilibrium states by
holding BECs for 8~s in a high magnetic field where $q_M=0$ and
$q_B/h=100$~Hz. The solid black line represents the MF ground
states (see text).} \label{100Hz}
\end{figure}

\begin{figure}[t]
\includegraphics[width=85mm]{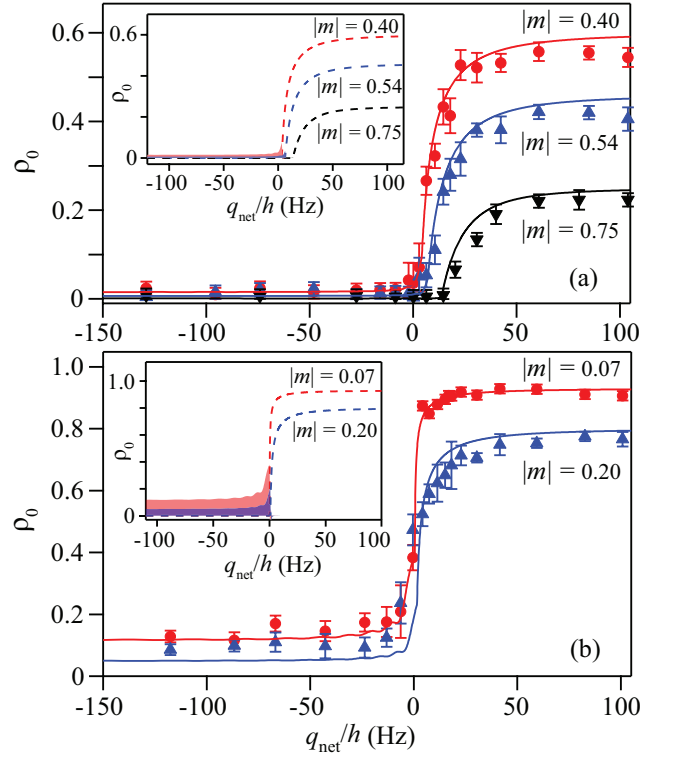}
\caption{(color online) $\rho_0$ as a function of $q_{\rm net}$ at
$t=t_f$ for three large $|m|$ in Panel (a) and for two small $|m|$
in Panel (b) in equilibrium states created by our new method.
Solid lines are simulation results for the experimental processes
based on Eq.~\ref{eqn:E2} (see text). Insets: dashed lines are the
MF ground states. Shaded areas represent the differences between
our simulation results and the MF theory at various $m$. The black, blue, and red colors in Panel (a) respectively correspond
to results at $|m|=$~0.75, 0.54, and 0.40. The blue and red colors
in Panel (b) represent results at $|m|=$~0.20 and 0.07,
respectively.} \label{largeM}
\end{figure}

\begin{figure*}[t]
\includegraphics[width=175mm]{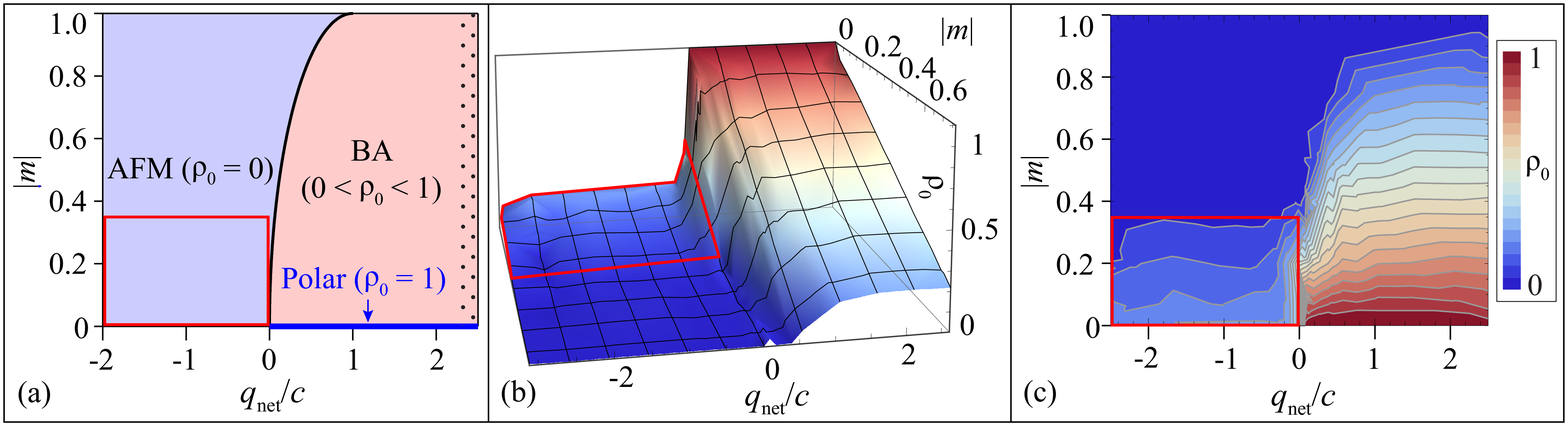}
\caption{(color online) (a). The MF phase diagram of spin-1
antiferromagnetic spinor BECs based on
Eqs.~(\ref{eqn:E1}-\ref{Eqn:Eg}). Our new method works everywhere
except in the area marked by red solid lines, while Method-O only
applies to the area filled with dots at large $q_{\rm net}$. Panel
(b) (or (c)) is a 3D (or a contour) plot of the experimental phase
diagram consisting of data taken by our new method at 153
different $q_{\rm net}$ and $m$. Red solid lines in Panels (b)-(c)
mark the region where our data are different from the MF ground
states.} \label{phase}
\end{figure*}

The SMA assumes all spin states share the same spatial
wavefunction, which has been a successful model to understand
spinor microcondensates~\cite{ZhaoUwave,
faraday,fluctuation,black,Gerbier2012,Chapman2005,You2005,Ueda2010,Lamacraft2011}.
After taking into account that $N$ and $m$ are independent of time
$t$ and neglecting all constant terms in the Hamiltonian of spinor
BECs, we use the SMA to express the BEC energy $E$ and the time evolution of
$\rho_0$ and $\theta$ as~\cite{StamperKurnRMP,You2005,Ueda2010}
\begin{align}  \label{eqn:E1}
E(t)=&~c\rho_0(t)\{[1-\rho_0(t)]+\sqrt{[1-\rho_0(t)]^2-m^2}\cos[\theta(t)]\}\nonumber\\
&+q_{\rm net}(t)[1-\rho_0(t)]~;\\
\dot{\rho_0}=&-\frac{4\pi}{h}\frac{\partial E(t)}{\partial \theta(t)},
~~\dot{\theta}=\frac{4\pi}{h}\frac{\partial E(t)}{\partial \rho_0(t)}~. \label{eqn:E2}
\end{align}
Here $q_{\rm net}=q_M+q_B$ is the net quadratic Zeeman energy with
$q_B$ (or $q_M$) being induced by magnetic (or microwave dressing)
fields. The spin-dependent interaction energy $c$ is proportional
to the atom density, and is positive (or negative) in $F$=1
antiferromagnetic $^{23}$Na (or ferromagnetic $^{87}$Rb) spinor
BECs. For example, $c/h$ is 40~Hz for our $^{23}$Na system in this paper,
where $h$ is the Planck constant. The fractional population
$\rho_{\rm m_F}$ and the phase $\theta_{\rm m_F}$ of each $m_F$
state are independent of position in SMA, and
$m=\rho_{+1}-\rho_{-1}$. The relative phase among the three $m_F$
spin states is $\theta=\theta_{+1} + \theta_{-1} -2\theta_0$.

By minimizing Eq.~(\ref{eqn:E1}), we find $\rho_0$ in a MF ground
state of $F$=1 spinor BECs is zero if $q_{\rm net} < \rm c(1 \pm
\sqrt{1-m^2})$; or equals to one if $m=0$ and $q_{\rm net}
> -c(1\pm 1)$; or is the root of the following equation at all other $q_{\rm net}$ and $m$,
\begin{equation}\label{Eqn:Eg}
c [1-2 \rho_0 \pm \frac{(1-2
\rho_0)(1-\rho_0)-m^2}{\sqrt{(1-\rho_0)^2-m^2}}]-q_{\rm net}=0~,
\end{equation}
where the $+$ (or $-$) sign applies to ferromagnetic (or
antiferromagnetic) spinor BECs. Typical MF ground states of spin-1
sodium BECs are shown in Figs.~\ref{100Hz} and \ref{largeM}. Our
experimental phase diagram and the theoretical phase diagram based
on Eq.~(\ref{eqn:E1}-\ref{Eqn:Eg}) are also plotted in
Fig.~\ref{phase}.

The experimental setup is similar to that elaborated in our recent
publications~\cite{JiangBEC,ZhaoUwave}. A $F$=1 BEC of $5\times
10^4$ atoms is created by a forced evaporation in a crossed
optical dipole trap. To fully polarize atoms into the $|F=1, m_F
=-1 \rangle$ state, we turn on a weak magnetic field gradient and
a low magnetic bias field in the forced evaporative cooling
process. A resonant rf-pulse of a proper amplitude and duration is
applied to prepare an initial state with any desired combination
of the three $m_F$ states. This moment is defined as the starting
point ($t=0$) of our experimental sequences, as shown in
Fig.~\ref{100Hz}(a). Every sequence ends at $t=t_f$. Populations
of multiple spin states are then measured by a standard
Stern-Gerlach absorption imaging.

We use two different methods to generate equilibrium states. The
Method-O is an old and widely-used method, which creates
equilibrium states simply by holding a BEC at a fixed $q_{\rm
net}$ for a sufficiently long time. We find that the required hold
time is longer than 2~s for all positive $q_{\rm net}$ studied in
this paper. This old method fails for our system in low magnetic
fields (i.e, the small positive $q_{\rm net}$ region), because
energy dissipates very slowly and the required hold time is longer
than the BEC lifetime ($\sim 10$~s) in this region. This old
method is more problematic in the negative $q_{\rm net}$ region,
because it leads to significant atom losses and detrimental
changes in $m$. In order to overcome these experimental challenges
associated with the old method, we have developed a new method,
Method-N. A comparison of these two methods starting from the same
initial state is shown in Fig.~\ref{100Hz}(b), which highlights
the advantage of our new method. A typical experimental sequence
of the new method is listed in Fig.~\ref{100Hz}(a). We first hold
a spinor BEC in the optical trap for 5~s at a very high magnetic
field with $q_B/h=900$~Hz. This step ensures the BEC reaches its
ground states, since we and Ref.~\cite{fluctuation} find that the
energy dissipation rate quickly increases with $q_B$. Second, we
adiabatically ramp the magnetic field down to $q_B/h=20$~Hz in
0.1~s, keep $q_B$ at this value for 0.3~s, and then turn on a far
off-resonant microwave pulse in 0.1~s. Third, we tune only the
frequency of this pulse slowly within 0.5~s, in order to
adiabatically sweep its corresponding microwave field to a desired
$q_{\rm net}$. Our approach to characterize microwave dressing
fields and the frequency tuning curve for adiabatically sweeping
$q_{\rm net}$ within the range of $-\infty$ to $+\infty$ are as
same as those illustrated in our previous work~\cite{ZhaoUwave}.

In theory, once a BEC is prepared into its ground state, the BEC
may stay in its ground state at each $q_{\rm net}$ when a microwave field is
adiabatically ramped~\cite{Duan2013}. We can thus initially check
whether the new method is applicable by comparing equilibrium
states created by both new and old methods in a region, $q_{\rm
net}\gg0$, where the old method has been proven to generate the MF
ground states~\cite{black,faraday,Gerbier2012,fluctuation}.
Figure~\ref{100Hz}(c) shows such comparisons made at $q_{\rm
net}(t=t_f)/h=100~\rm{Hz}$ for various magnetization $m$. The
equilibrium states created by the two methods appear to be quite
similar, and they stay very close to the same black solid line
which represents the MF ground states in Fig.~\ref{100Hz}(c). This
suggests that our new method is adiabatic enough to replace the
old method in studies related to the BEC phase diagrams.
We also find that a spinor BEC returns to its original state when
we ramp a microwave field from $q_M=0$ to a fixed nonzero $q_M$
and then back to $q_M=0$ with this new method, although this
observation may not be sufficient to prove the process is
adiabatic.

We then apply our new method to a much wider range of $q_{\rm
net}$ and $m$, especially in the negative and small positive
$q_{\rm net}$ regions which cannot be easily explored by the old
method, as shown in Fig.~\ref{largeM}. We find two interesting
results from this figure. First, our data in Fig.~\ref{largeM}(a)
show a quantum phase transition between a BA phase and an AFM
phase at each $m$. This BA-AFM phase transition appears to occur
at a larger $q_{\rm net}$ when $|m|$ gets bigger, which can be
well explained by the MF theory (i.e., dashed lines in the inset
in Fig.~\ref{largeM}(a)). Another interesting result is that this
new method does allow us to access many previously unexplored
regions in the phase diagram, although there is a visible
discrepancy between the MF ground states and our data at a small
$m$ in the negative $q_{\rm net}$ region, as shown in
Fig.~\ref{largeM}(b). To understand this phenomenon, we simulate
the experimental processes based on Eq.~\ref{eqn:E2} by taking a
proper formula to account for the time evolution of $q_{\rm net}$
during an adiabatic ramping of microwave fields.
Figure~\ref{largeM} shows that the simulation results can well
resemble the experimental data, while the differences between our
simulation results and the MF ground states are emphasized by a
shaded area at each $m$ in the two insets in Fig.~\ref{largeM}.
These shaded areas appear to slowly increase in the negative
$q_{\rm net}$ region when $|m|$ approaches zero. In other words,
the discrepancy between our data and the MF ground states only
becomes noticeable at a small $|m|$ in the negative $q_{\rm net}$
region. Due to this discrepancy, we find that the predicted
quantum phase transition between an AFM phase and a longitudinal
polar phase at $m=0$ and $q_{\rm net}=0$ is replaced by a
transition between a BA phase and a longitudinal polar phase.
Since our experimental resolution for $\rho_0$ is around 0.02,
Fig.~\ref{largeM} implies that our new method is sufficient to map
out the BEC phase diagram in the positive $q_{\rm net}$ region at
each $m$, and in the negative $q_{\rm net}$ as long as $|m|\geq
0.4$.

\begin{figure}[t]
\includegraphics[width=85mm]{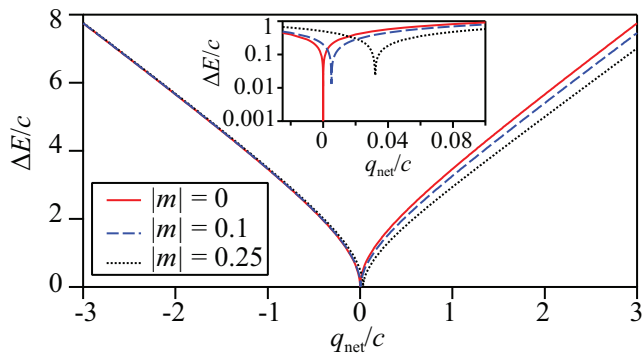}
\caption{(color online) The energy gap $\Delta E$ in the unit of
$c$ as a function of $q_{\rm net}/c$ at three $|m|$ based on
Eq.~\ref{eqn:H} (see text). Inset: a magnified view of the main
figure in the region of $-0.025<q_{\rm net}/c<0.1$.}
\label{rhoVSm}
\end{figure}
Figure~\ref{phase} clearly summarizes the improvement provided by
this new method, after comparing the theoretical MF phase diagram
to an experimental phase diagram consisting of our data taken at
153 different $q_{\rm net}$ and $m$. All three predicted phases
(i.e., an AFM, a polar, and a BA phases), an AFM-BA phase
transition at a fixed $m$, and a transition between a longitudinal
polar phase and a BA phase at a certain positive $q_{\rm net}$ are
shown in the experimental phase diagram. Good agreements between
our data and the MF ground states can be found everywhere in the
two phase diagrams except in the region where $|m|<0.4$ and
$q_{\rm net}<0$. This problematic region has been marked by red
solid lines in Fig.~\ref{phase}. Ramping microwave fields at a
slower rate should help to diminish this problematic region,
however, a slower rate requires holding a BEC in microwave fields
for a longer time and thus inevitably leads to more atom losses
and a bigger change in $m$. In fact, we tried quite a few
different microwave ramping rates, but none of them enabled a
spinor BEC to reach its MF ground states when $m$ is very small
and $q_{\rm net}<0$. The same problem also exists in simulation
results: our simulation program cannot suggest a reasonable
ramping rate to ensure an adiabatic sweep of $q_{\rm net}$ across
a phase transition for a small $m$.

To understand this problem, we need to find the exact value of
$\Delta E$, the energy gap between the ground state and the first
excited state in spinor BECs. Similarly to Ref.~\cite{Duan2013},
we can describe a spinor BEC in the Fock space. The spin-dependent
part of the Hamiltonian in a $F$=1 spinor BEC can be expressed
as~\cite{Duan2013,You2005,You2003}
\begin{equation} \label{eqn:H}
H=\sum_{i,j,k,l=-1}^{1}\left [q_{\rm net} k^2 a_k^\dagger
a_k+\frac{c}{2} \sum_{\gamma}a_k^\dagger
a_i^\dagger(F_\gamma)_{ij}(F_\gamma)_{kl} a_j a_l\right ],
\end{equation}
since $m$ is conserved and there are only a finite number of atoms
in a typical equilibrium state studied in this paper. Here $a_k$
($a_k^\dagger$) is the annihilation (creation) operator of the
$|F=1, m_F=k\rangle$ state, and $F_{\gamma=x,y,z}$ are the spin-1
matrices. By diagonalizing the Hamiltonian in Eq.~\ref{eqn:H} and
performing an exact numerical many-body calculation, we can find
the energy gaps. Figure~\ref{rhoVSm} shows numerical examples of
$\Delta E$ at three typical $|m|$. It appears that $\Delta E$
drastically drops by more than three orders of magnitude when
$|m|$ and $q_{\rm net}$ approach zero, as shown in the inset in
Fig.~\ref{rhoVSm}. Therefore, it is not surprising that
adiabatically sweeping $q_{\rm net}$ across a quantum phase
transition point is not feasible at a very small $m$, especially
at $m=0$.

In conclusion, we have observed two types of quantum phase
transitions in a spin-1 antiferromagnetic spinor BEC, and
developed a new method to create the equilibrium states of spinor
condensates by adiabatically sweeping a microwave field. The
biggest advantage of this method is to avoid significant atom
losses and detrimental changes in $m$ at large microwave fields.
We have demonstrated that this method is able to map out the phase
diagram of antiferromagnetic spinor BECs for all $m$ in the
positive $q_{\rm net}$ region and for all negative $q_{\rm net}$
as long as $|m|\geq 0.4$. This method can be applied to other
atomic species when applying microwave fields are required. In
addition, adiabatically sweeping $q_{\rm net}$ across a quantum
phase transition demonstrated in this paper may be a big step towards confirming other important predictions, for
example, realizing massive entanglement in the vicinity of the Dicke
states with spinor BECs~\cite{Duan2013}.

We thank the Army Research Office, the National Science
Foundation, and the Oklahoma Center for the Advancement of Science
and Technology for financial support. M.W. thanks the Niblack
Research Scholar program.


\begin{thebibliography}{24}%
\makeatletter
\providecommand \@ifxundefined [1]{%
 \@ifx{#1\undefined}
}%
\providecommand \@ifnum [1]{%
 \ifnum #1\expandafter \@firstoftwo
 \else \expandafter \@secondoftwo
 \fi
}%
\providecommand \@ifx [1]{%
 \ifx #1\expandafter \@firstoftwo
 \else \expandafter \@secondoftwo
 \fi
}%
\providecommand \natexlab [1]{#1}%
\providecommand \enquote  [1]{``#1''}%
\providecommand \bibnamefont  [1]{#1}%
\providecommand \bibfnamefont [1]{#1}%
\providecommand \citenamefont [1]{#1}%
\providecommand \href@noop [0]{\@secondoftwo}%
\providecommand \href [0]{\begingroup \@sanitize@url \@href}%
\providecommand \@href[1]{\@@startlink{#1}\@@href}%
\providecommand \@@href[1]{\endgroup#1\@@endlink}%
\providecommand \@sanitize@url [0]{\catcode `\\12\catcode
`\$12\catcode
  `\&12\catcode `\#12\catcode `\^12\catcode `\_12\catcode `\%12\relax}%
\providecommand \@@startlink[1]{}%
\providecommand \@@endlink[0]{}%
\providecommand \url  [0]{\begingroup\@sanitize@url \@url }%
\providecommand \@url [1]{\endgroup\@href {#1}{\urlprefix }}%
\providecommand \urlprefix  [0]{URL }%
\providecommand \Eprint [0]{\href }%
\providecommand \doibase [0]{http://dx.doi.org/}%
\providecommand \selectlanguage [0]{\@gobble}%
\providecommand \bibinfo  [0]{\@secondoftwo}%
\providecommand \bibfield  [0]{\@secondoftwo}%
\providecommand \translation [1]{[#1]}%
\providecommand \BibitemOpen [0]{}%
\providecommand \bibitemStop [0]{}%
\providecommand \bibitemNoStop [0]{.\EOS\space}%
\providecommand \EOS [0]{\spacefactor3000\relax}%
\providecommand \BibitemShut  [1]{\csname bibitem#1\endcsname}%
\let\auto@bib@innerbib\@empty
\bibitem{StamperKurnRMP} D. M. Stamper-Kurn and M. Ueda, \bibinfo{journal}{Rev. Mod. Phys.} \textbf{\bibinfo{volume}{85}}, \bibinfo{pages}{1191} (\bibinfo{year}{2013}).

\bibitem{Ueda} Y. Kawaguchi and M. Ueda, \bibinfo{journal}{Phys. Rep.} \textbf{\bibinfo{volume}{520}}, \bibinfo{pages}{253} (\bibinfo{year}{2012}).

\bibitem{Duan2013} Z. Zhang and L.-M. Duan, \bibinfo{journal}{Phys. Rev. Lett.} \textbf{\bibinfo{volume}{111}}, \bibinfo{pages}{180401} (\bibinfo{year}{2013}).

\bibitem{Chapman2013} T. M. Hoang, C. S. Gerving, B. J. Land, M. Anquez, C. D. Hamley, and M. S. Chapman, \bibinfo{journal}{Phys. Rev. Lett.}
  \textbf{\bibinfo{volume}{111}}, \bibinfo{pages}{090403} (\bibinfo{year}{2013}).

\bibitem{Chapman2012} C. D. Hamley, C. S. Gerving, T. M. Hoang, E. M. Bookjans, and M. S. Chapman, Nature Physics \textbf{8}, 305
(2012).

\bibitem{JiangBEC} J. Jiang, L. Zhao, M. Webb, N. Jiang, H. Yang, and Y. Liu, \bibinfo{journal}{Phys. Rev. A} \textbf{\bibinfo{volume}{88}}, \bibinfo{pages}{033620} (\bibinfo{year}{2013}).

\bibitem{Raman2011} E. M. Bookjans, A. Vinit, and C. Raman, \bibinfo{journal}{Phys. Rev. Lett.} \textbf{\bibinfo{volume}{107}}, \bibinfo{pages}{195306} (\bibinfo{year}{2011}).

\bibitem{black} A. T. Black, E. Gomez, L. D. Turner, S. Jung, and P. D. Lett, \bibinfo{journal}{Phys. Rev. Lett.}
  \textbf{\bibinfo{volume}{99}}, \bibinfo{pages}{070403} (\bibinfo{year}{2007}).

\bibitem{faraday} Y. Liu, S. Jung, S. E. Maxwell, L. D. Turner, E. Tiesinga, and P. D. Lett, \bibinfo{journal}{Phys. Rev. Lett.} \textbf{\bibinfo{volume}{102}}, \bibinfo{pages}{125301} (\bibinfo{year}{2009}).

\bibitem{fluctuation} Y. Liu, E. Gomez, S. E. Maxwell, L. D. Turner, E. Tiesinga, and P. D. Lett, \bibinfo{journal}{Phys. Rev. Lett.} \textbf{\bibinfo{volume}{102}}, \bibinfo{pages}{225301} (\bibinfo{year}{2009}).

\bibitem{Gerbier2012} D. Jacob, L. Shao, V. Corre, T. Zibold, L. De Sarlo, E. Mimoun, J. Dalibard, and F. Gerbier, \bibinfo{journal}{Phys. Rev. A} \textbf{\bibinfo{volume}{86}}, \bibinfo{pages}{061601} (\bibinfo{year}{2012}).

\bibitem{ZhaoUwave} L. Zhao, J. Jiang, T. Tang, M. Webb, and Y. Liu, \bibinfo{journal}{Phys. Rev. A} \textbf{\bibinfo{volume}{89}}, \bibinfo{pages}{023608} (\bibinfo{year}{2014}).

\bibitem{Chapman2005} M.-S. Chang, Q. Qin, W. Zhang, L. You, and M. S. Chapman, \bibinfo{journal}{Nature Physics}
  \textbf{\bibinfo{volume}{1}}, \bibinfo{pages}{111} (\bibinfo{year}{2005}).

\bibitem{bloch} A. Widera, F. Gerbier, S. F\"{o}lling, T. Gericke, O. Mandel, and I. Bloch, \bibinfo{journal}{New Journal of Physics} \textbf{\bibinfo{volume}{8}}, \bibinfo{pages}{152} (\bibinfo{year}{2006}).

\bibitem{f2sengstock2} J. Kronj\"{a}ger, C. Becker, P. Navez, K. Bongs, and K. Sengstock, \bibinfo{journal}{Phys. Rev. Lett.} \textbf{\bibinfo{volume}{97}}, \bibinfo{pages}{110404} (\bibinfo{year}{2006}).

\bibitem{f2sengstock1} H. Schmaljohann, M. Erhard, J. Kronj\"{a}ger, M. Kottke, S. van Staa, L. Cacciapuoti, J. J. Arlt, K. Bongs, and K. Sengstock, \bibinfo{journal}{Phys. Rev. Lett.} \textbf{\bibinfo{volume}{92}}, \bibinfo{pages}{040402} (\bibinfo{year}{2004}).

\bibitem{f2Hirano} T. Kuwamoto, K. Araki, T. Eno, and T. Hirano, \bibinfo{journal}{Phys. Rev. A}
  \textbf{\bibinfo{volume}{69}}, \bibinfo{pages}{063604} (\bibinfo{year}{2004}).

\bibitem{uwaveBloch} F. Gerbier, A. Widera, S. F\"{o}lling, O. Mandel, and I. Bloch, \bibinfo{journal}{Phys. Rev. A} \textbf{\bibinfo{volume}{73}}, \bibinfo{pages}{041602(R)} (\bibinfo{year}{2006}).

\bibitem{Stamper-Kurn2009} S. R. Leslie, J. Guzman, M. Vengalattore, J. D. Sau, M. L. Cohen, and D. M. Stamper-Kurn, \bibinfo{journal}{Phys. Rev. A} \textbf{\bibinfo{volume}{79}}, \bibinfo{pages}{043631} (\bibinfo{year}{2009}).


\bibitem{You2005} W. Zhang, D. L. Zhou, M.-S. Chang, M. S. Chapman, and L. You, \bibinfo{journal}{Phys. Rev. A} \textbf{\bibinfo{volume}{72}}, \bibinfo{pages}{013602} (\bibinfo{year}{2005}).

\bibitem{Ueda2010} Y. Kawaguchi, H. Saito, K. Kudo, and M. Ueda, \bibinfo{journal}{Phys. Rev. A} \textbf{\bibinfo{volume}{82}}, \bibinfo{pages}{043627} (\bibinfo{year}{2010}).

\bibitem{Lamacraft2011} A. Lamacraft, \bibinfo{journal}{Phys. Rev. A} \textbf{\bibinfo{volume}{83}}, \bibinfo{pages}{033605} (\bibinfo{year}{2011}).

\bibitem{You2003} W. Zhang, S. Yi, and L. You, \bibinfo{journal}{New Journal of Physics} \textbf{\bibinfo{volume}{5}}, \bibinfo{pages}{77} (\bibinfo{year}{2003})

\end{thebibliography}
\end{document}